\documentclass[aps,twocolumn,superscriptaddress,showpacs,prl,floatfix]{revtex4-1}

\usepackage{times}
\usepackage{braket}
\usepackage{amsmath,txfonts}
\usepackage[dvips,dvipdfmx]{graphicx}
\usepackage{wrapfig}
\usepackage{color}

\begin{document}
\author{Emika Takata}
\affiliation{Institute for Solid State Physics, University of Tokyo, Kashiwa 277-8581, Japan}
\author{Tsutomu Momoi}
\affiliation{Condensed Matter Theory Laboratory, RIKEN, Wako, Saitama 351-0198, Japan}
\affiliation{RIKEN Center for Emergent Matter Science (CEMS), Wako, Saitama, 351-0198, Japan}
\author{Masaki Oshikawa}
\affiliation{Institute for Solid State Physics, University of Tokyo, Kashiwa 277-8581, Japan}
\title{Nematic ordering in pyrochlore antiferromagnets: high-field phase of chromium spinel oxides}
\date{\today}
\begin{abstract}
Motivated by recent observation of a new high field phase near saturation in chromium spinels
$A$Cr$_2$O$_4$ ($A=$ Zn, Cd, Hg),
we study the $S = 3/2$ pyrochlore Heisenberg antiferromagnet with biquadratic interactions.
Magnon instability analysis at the saturation field reveals that a very small biquadratic interaction
can induce magnon pairing in pyrochlore antiferromagnets, which leads to the emergence of
a ferro-quadrupolar phase, or equivalently a spin nematic phase, below the saturation field.
We present the magnetic phase diagram in an applied field, studying both $S=3/2$ and $S=1$ spin systems.
The relevance of our result to chromium spinels is discussed.
\end{abstract}

\maketitle

Highly frustrated magnets have been an active playground to
find novel states of matter.
Frustration suppresses long-range magnetic orders, leading to
the possibility of quantum spin liquids~\cite{balents} without any
conventional order, or of other exotic states with unconventional
orders.
Among the latter, quantum multipolar states are of a particular interest.
The simplest among them, the quadrupolar state is also known as
the spin nematic state.
This state does not have any conventional magnetic order
with the spin vector as an order parameter.
Instead, it has a directional order whose order parameters are
the symmetric rank-2 spin tensors \cite{blume_hsieh,andreev}.
Quantum mechanically, such a state can be understood as a result
of condensation of bound magnon pairs.
The presence of such an exotic state is theoretically proposed in
various spin systems, including frustrated
ferromagnets~\cite{Chubukov,ShannonMS,MomoiSK}
and the one-dimensional $S=1/2$ zigzag chain compound
LiCuVO$_4$ \cite{hikihara_licuvo,zhitomirsky_licuvo}.
The emergence of multipolar phases has been also theoretically well
established in $S=1$ spin systems with bilinear and biquadratic interactions
on some lattices, such as square and cubic lattices,
when the biquadratic interactions are strong
enough~\cite{blume_hsieh,tanaka2001,harada}.
Despite these theoretical results,
the experimental confirmation of the spin nematic phase
still remains difficult~\cite{nawa}.

The nearest-neighbor Heisenberg antiferromagnet on the pyrochlore lattice
is one of the most 
frustrated spin systems.
The ground states are infinitely degenerate in the classical limit
\cite{reimers} and it is believed that there is no spin long-range order
at any temperature both in the classical ($S\rightarrow \infty$)
\cite{moessner} and quantum ($S=1/2$) \cite{canals} models. In real
materials this massive degeneracy can be often lifted by further
neighbor interactions or by the \emph{order-by-distortion}
effect \cite{yamashita,tchernyshyovA,tchernyshyovB,penc_pyrochlore}.

The chromium spinel oxides $ A \mathrm{Cr}_2 \mathrm{O}_4 $
($A=\mathrm{Hg},\mathrm{Cd},\mathrm{Zn} $) are ideal $S=3/2$ pyrochlore
antiferromagnets, where detailed comparison between theoretical and
experimental results is possible.
They have isotropic antiferromagnetic spin interactions, with a significant
spin-lattice coupling.
The purpose of the present Letter is to show that, in
the $S=3/2$ pyrochlore antiferromagnet under a strong magnetic
field close to saturation, a spin nematic phase is stabilized
by the spin-lattice coupling but without any crystal distortion.
We identify the novel phase discovered experimentally
in chromium spinel oxides near the saturation field
with the spin nematic (quadrupolar) phase.
Thus, the chromium spinel oxides hopefully have realized the
long-sought but so far experimentally elusive spin nematic
phase.

In all of the chromium spinel oxides mentioned above,
several ordered phases including a wide 1/2-magnetization plateau phase
appear at very low temperatures \cite{Cd_earlier,Hg_mag,Zn_mag},
in a magnetic field.
At the transition temperatures, both the lattice distortion and the magnetic ordering appear
simultaneously with lowering temperature \cite{Zn:spin-Peiers}.
Penc \textit{et al}.~\cite{penc_pyrochlore} theoretically analyzed these magnetic behaviors considering the coupling
between spin and lattice vibration degrees of freedom \cite{tchernyshyovA,tchernyshyovB}.
Integrating out the lattice degrees of freedom, they derived the following effective spin Hamiltonian, which has both bilinear
and biquadratic spin interactions,
\begin{equation}
  \mathcal{H} = \sum_{\langle i,j \rangle} \left[J_1 \mathbf{S}_i \cdot \mathbf{S}_j + J_2 (\mathbf{S}_i \cdot \mathbf{S}_j)^2 \right]-h \sum_{i} S_i^z,
\label{eq:j1-j2}
\end{equation}
where the bilinear exchange $ J_1 $ is positive (antiferromagnetic), the biquadratic exchange $ J_2 $ is negative,
$ \mathbf{S}_i $ is the $ S=3/2 $ spin operator on the $ i $-site, $ h $ is the magnetic field, and
the first summation is taken over all the nearest-neighbor pairs.
Analyzing this model in the classical ($S\rightarrow\infty$)
limit with a mean-field (MF) ansatz~\cite{penc_pyrochlore},
they showed a stable wide half-magnetization plateau
in a magnetic field
\footnote{
In fact, the ground state of the classical
Hamiltonian~\protect\eqref{eq:j1-j2}
has a macroscopic degeneracy as in the original
pyrochlore Heisenberg antiferromagnet~\cite{reimers,moessner}.
The macroscopic degeneracy would be lifted by a weak inter-tetrahedral
interaction, such as the third-neighbor ferromagnetic exchange~\cite{motome},
resulting in a ground state which may match with
that witin the mean-field ansatz~\cite{penc_pyrochlore}.
For simplicity, in this Letter, we implicitly assume the presence of such
a weak interaction.
We emphasize that, the spin nematic phase, which is our main focus
in this Letter, is stable without such an interaction; our conclusion
thus stands as long as the inter-tetrahedral interaction
is sufficiently weak.
}.
They also found canted antiferromagnetic
ordered phases in the magnetic field above and below the
field-range of the plateau phase.

Later, this effective model was further extended to spin
 models with generalized four-spin exchange interactions
 by a precise symmetry analysis of phonons \cite{bpmodel} and
 a consideration of local site distortions (the Einstein model) \cite{emodel}.
 These extended models also showed the half-magnetization plateau phase
 with the same spin ordered structure as the model (\ref{eq:j1-j2}), in the classical limit.

The magnetic structure of antiferromagnetic phases
derived from the model (\ref{eq:j1-j2}) in the classical limit
with the MF ansatz were found to be consistent with the available
experimental results by then \cite{Hg_ESR}.
However, the recently developed high-field measurements discovered the presence of a classically unexpected
phase just below the fully polarized (FP) phase and above the canted
antiferromagnetic and plateau phases,
in all the three compounds \cite{Cd_new, Hg_ESR, Zn}.
ESR measurements reveal that the lattice distortion is trigonal in the spin canted phase.
In contrast, in the phase found out newly,
the distortion is relaxed
although the magnetization is not still saturated~\cite{Hg_ESR}.
Magneto-optical measurements also implies that the lattice has a high
symmetry and spin excitations with $\Delta S^z=1$ are suppressed in this
new phase \cite{Hg_MO}.

\begin{figure}[t]
\begin{center}
  \includegraphics[scale=0.5]{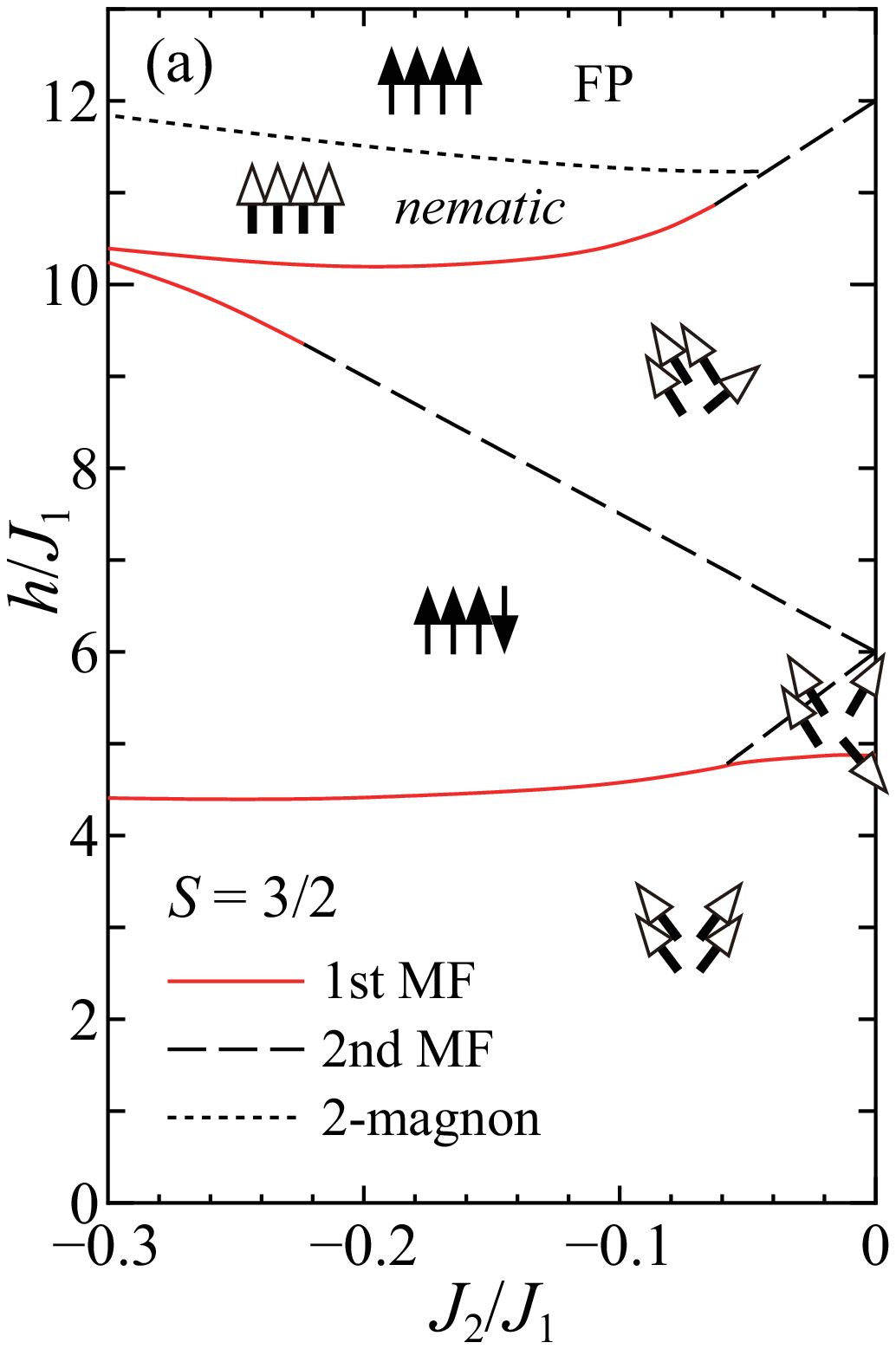}\hspace*{2mm}
  \mbox{\raisebox{12mm}{\includegraphics[width=30mm]{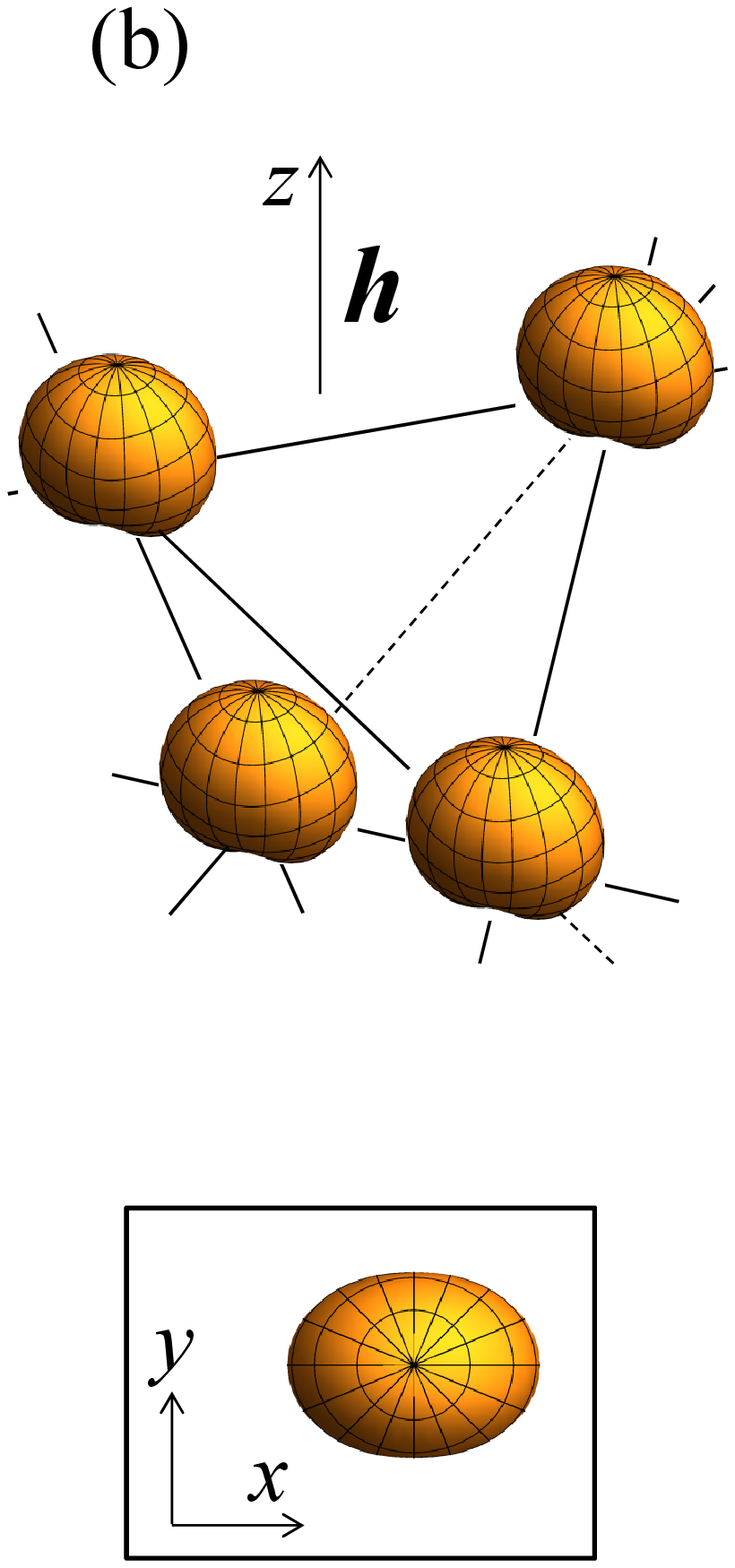}}}
  \caption{(Color online) (a) Magnetic phase diagram of the $S=3/2$ pyrochlore antiferromagnet with the
 biquadratic interaction in a magnetic field $h$.
Solid (dashed)
lines denote 1st (2nd) order phase boundaries in the mean-field (MF) approximation.
Filled arrows represent
fully polarized (FP) magnetic moments and empty arrows partially polarized
ones.
The dotted line shows the exact two-magnon instability line.
The partially polarized uniform phase below this dotted line is a spin nematic phase.
  (b) Schematic figure of $S=3/2$ spins in the spin nematic phase.
  The overlap probability between the partially polarized $S=3/2$ quadrupolar moment
  and the coherent spin state \cite{radcliffe} is plotted on each site.
  Inset: top view of the quadrupolar moment showing the spin anisotropy in the $xy$ plane.
  }
  \label{fig:pd}
\end{center}
\end{figure}

In this Letter, motivated by the experimental discovery of the novel high-field phase,
we study quantum effects
in the $S=3/2$ pyrochlore antiferromagnet~(\ref{eq:j1-j2})
with biquadratic interaction
in an applied field.
In fact, not much is known about multipolar order in $S>1$ spin systems.
In a MF approximation study of
a $S=3/2$ spin system on the cubic lattice at zero field,
biquadratic interactions did not induce any multipolar order~\cite{fridman}.
However, the possibility of the spin nematic phase for $S=3/2$
has to be examined more carefully beyond these limitations, especially in the
light of the experimental results on chromium spinels near the
saturation field.

We first employ the site-decoupled MF 
approximation.
The obtained magnetic
phase diagram in this approximation is almost the same as the classical
phase diagram.
However, it reveals that quantum multipolar states exist
close to the lowest energy state near the saturation field,
and in fact become degenerate with the lowest energy state
at the MF 
phase boundary.
This suggests the possibility of an emergence of multipolar states
near the saturation field, in more precise analysis.
Indeed, in a magnon-instability analysis on the FP
state, which is exact just below the saturation field,
we find an emergence of a spin nematic phase.
[See Fig.~\ref{fig:pd}(a)].

The site-decoupled MF 
approximation is a variational method
within direct-product states of on-site quantum states
$ \Ket{\Psi}=\bigotimes_i \Ket{\phi}_i $.
The arbitrary local states 
can be expressed by the linear combination of {$ \Ket{m} $}, where 
$ \Ket{m} $ is the eigenstate for the spin $ z $ component $ S^z $ on a single site with eigenvalue $m$.
For spin 3/2, which is of our main interest in this Letter,
the product states $ \Ket{\Psi} $ do include
the nematic state
$ \bigotimes_i \left( \Ket{\frac{3}{2}}_i+e^{i\gamma} \Ket{-\frac{3}{2}}_i\right) $ \cite{chubukov1990,fridman}.
Fixing the overall phase and normalization, one can describe
the local spin-3/2 state $\Ket{\phi}_i $ with 6 real parameters.

Furthermore, we assume that the system has 4-sublattice structure, $
\Ket{\Psi}=\bigotimes_{\alpha} \bigotimes_{i \in \alpha}
\Ket{\phi_{\alpha}}_i $, where $ \alpha $ labels the sublattices and
$i$-site belongs to the $ \alpha $-sublattice. In total, there are $ 6
\times 4=24 $ real variational parameters.  Minimizing the energy
expectation value $ \Braket{\Psi|\mathcal{H}|\Psi} $, we obtain the
phase diagram of antiferromagnetic phases 
in Fig.~\ref{fig:pd}(a).
The black dashed lines and the red solid lines, respectively, denote the
second order and first order phase boundaries.

The magnetic characteristics of these antiferromagnetic phases
are essentially  the same as in the classical limit \cite{penc_pyrochlore}.
In the lower field regime, there appears canted antiferromagnetic phase 
with 2:2-sublattice structure.
This state changes to higher field phases through first order phase transition.
In higher field, the up-up-up-down plateau phase 
appears in a wide field range.
When $J_2$ is relatively weak, this state continuously
changes to a canted antiferromagnetic phase 
with 3:1 sublattice structure with increasing field
and also to another canted antiferromagnetic phase 
with 2:1:1-sublattice structure with decreasing field.
The upper boundary of the plateau phase expands to the higher field regime in comparison with
the MF 
phase diagram in the classical spin model \cite{penc_pyrochlore}.
This is due to quantum effects originated from the biquadratic interaction.
For strong $|J_2|$ regime, the plateau phase jumps to the FP phase through the first order
phase transition with increasing field.

We should also note that multipolar states are also stable solutions in
a finite magnetic field in the MF 
approximation under the
condition that all sites are equivalent.  These multipolar MF 
solutions are highly degenerate.  Both 
purely quantum states such as a quadrupolar state $ \bigotimes_i \left(a
\Ket{\frac{3}{2}}+c \Ket{-\frac{1}{2}} \right) $ and a octupolar state $
\bigotimes_i \left(a^\prime \Ket{\frac{3}{2}}+c^\prime
\Ket{-\frac{3}{2}} \right) $ and also a continuous deformation of these
two states belong to the degenerate manifold of the MF 
solutions.
In the site-decoupled MF 
approximation, these states generally
have higher energy than the four-sublattice canted states and
half-magnetization plateau state.  However the MF 
energies of
the degenerate multipolar states and the plateau state become identical
exactly at the phase boundary between the plateau phase and the FP
phase.  This coincidence of the MF 
energies at the saturation field
can open a possibility of emergence of multipolar phases above the
plateau phase in a more precise analysis.

To investigate this possibility,
we study instability induced by
magnon excitations in the FP phase.
Above the saturation field, all spins are perfectly polarized along the
magnetic field.
Below the saturation field,
magnons, which are flipped spins in the FP state,
are sparsely induced and their condensation leads to
ordering of magnetic structure transverse to the applied field.
Bose-Einstein condensation (BEC) of single magnons realizes the spin canted
antiferromagnetic states \cite{matsubara}.
The nature of the condensed magnons determine the magnetic structure.

For the single-magnon excitations in the FP state, we obtain 
four branches in the excitation energy spectrum:
$ E^{\pm} (\mathbf{k}) $ and $ E^0 $, where $ E^0 $ is doubly degenerate and independent on the wave vector i.e.\ it has a flat mode.
The lowest excitation mode is $ E^0 $ for $ 2 J_1+3 J_2 \ge 0 $
and $ E^-(\mathbf{k}) $ for $ 2 J_1+3J_2 <0 $.
The saturation field $ h_{s1} $ given by the one-magnon instability is
$ h_{s1} = 6(2J_1+3J_2)$
in the parameter region $2 J_1+3J_2 \ge 0$,
whereas the ferromagnetic ground state is stable against single magnon
excitations at zero field in $2 J_1+3J_2 <0$.
In the pyrochlore antiferromagnetic Heisenberg model with $J_2=0$,
it is known that the lowest-energy magnon states
are localized~\cite{magnon_localize}.
This is also true for a finite $J_2$.

Ground states of dilute magnons are highly degenerate,
since the single-magnon eigenstates are localized.
In this case,
even a weak perturbation can change the nature of the ground state
drastically.  Actually, the biquadratic interaction has a pair hopping
process $(S^+_i)^2 (S^-_j)^2 + h.c.$,
which can induce a dynamical process of two magnons.
A similar dynamical energy gain happens to two magnons in
frustrated ferromagnets \cite{ShannonMS} where two-magnons become
stable bound states and these bound magnon pairs condense in the  ground state
\cite{Chubukov,ShannonMS,MomoiSK,hikihara_licuvo,zhitomirsky_licuvo}.
Such an unconventional magnon condensed state is nothing but
the spin nematic state, where
the order parameter is given by quadrupolar operators
$(S^x_i S^x_j-S^y_i S^y_j,S^x_i S^y_j+S^y_i S^x_j)$
defined on single sites $(i=j)$
or on neighboring sites.
To analyze the emergence of bound-magnon condensation below the
saturation field, we study multiple-magnon instability to the FP state
at the saturation field in the following.

We first consider the two-magnon excited states.  Two-magnon excited
states are generally expressed as $ \Ket{2}=\sum_{(i,j)} \phi_{ij} S^-_i
S^-_j \Ket{0} $, where the summation is taken over all pairs of lattice
sites.
We solve the Schr\"odinger equation for this wave function exactly,
extending the previous method~\cite{mattis}
to a system with plural sites in the unit cells.
The energy spectrum contains a branch of stable two magnon bound state,
which has a lower energy than two independent localized magnons at the
saturation field
for $J_2<- 0.043 J_1$.
The dynamical process effectively gives the attractive interaction between two magnons.
The lowest excitation mode is dispersive and non-degenerate, and has the minimum energy
at $ \mathbf{k}=(0,0,0) $,
where $ \mathbf{k} $ is the center-of-mass wave vector.
It means that the magnon pairs
can move around on the lattice.
From the lowest excitation energy,
we obtain the saturation field $ h_{s2} $ at which the lowest two-magnon bound state becomes gapless.
The values of $ h_{s2} $ are plotted in Fig.\ \ref{fig:pd}(a).
For $J_2<- 0.043 J_1$, $ h_{s2} $ is larger than both $ h_{s1} $ and the upper boundary field of 1/2-plateau
phase. We thus find that a nematic phase is expected to
appear below $ h_{s2} $ induced by the two-magnon instability.

\begin{figure}
\begin{center}
  \includegraphics[width=6cm,clip]{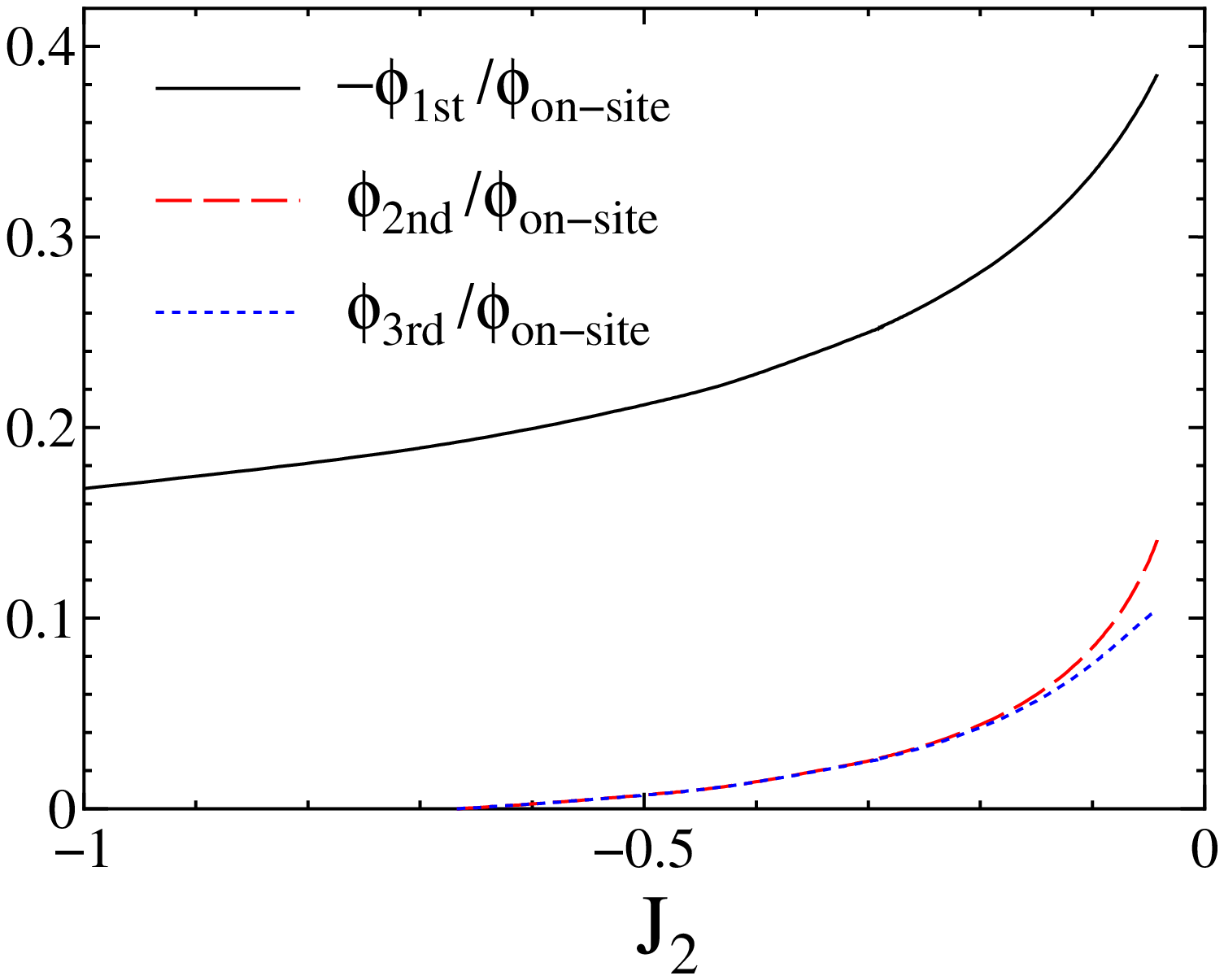}
  \caption{(Color online) Relative position dependence in the wave function of
  the lowest two-magnon bound state.}
  \label{fig:gs}
\end{center}
\end{figure}

We can identify the nature of the quadrupolar order from the structure of the magnon pair.
The wave-function of the lowest-energy bound magnon pair
is uniform, i.e.\ it has the wave vector $ \mathbf{k}=(0,0,0) $,
and fully symmetric ($ A_1 $) under the space symmetry group
$O_h$.
Hence that is simply represented in the form
$ \Ket{b}=\phi_{\mathrm{on-site}} \sum_{i} S_i^- S_i^- \Ket{0} + \phi_{\mathrm{1st}} \sum_{\Braket{i,j}_1} S_i^- S_j^- \Ket{0} + \phi_{\mathrm{2nd}} \sum_{\Braket{i,j}_2} S_i^-S_j^- \Ket{0} + \cdots $, where $ \sum_{\Braket{i,j}_n} $ is taken over the  $ n $-th neighboring pairs and the coefficients $ \phi $
are equal for all the pairs in each type.
We show the ratios of these coefficients $ \phi $ in Fig. \ref{fig:gs}.
The two bound magnons 
are almost confined to the same site or to the nearest neighbor sites
for large $J_2$.
However, for small $|J_2|$ the two magnons are more weakly bound,
occupying farther sites, and the size of the pair diverges as $J_2 \to 0$.
We note that, if we restrict the Hilbert space of magnon pairs to
on-site pairs, the estimated saturation field reduces to that
from the MF 
approximation.  Thus the correction to the MF
approximation originates from the fact that the bound magnon
pairs are not totally confined on single sites.

The condensate of these magnon pairs
is characterized by ferro-quadrupolar order
$\langle S_j^- S_j^- \rangle=\sqrt{\rho_0} e^{2i\theta}$ on the single sites
[see Fig.\ \ref{fig:pd}(b)]
and also $\langle S_j^- S_k^- \rangle=- \sqrt{\rho_1} e^{2i\theta}$ on
the nearest-neighbor bonds $(j,k)$.
We note that the stability of the bi-magnon condensate depends on the
correlation effects between magnon pairs, which remain to be studied.

\begin{figure}[t]
\begin{center}
    \includegraphics[scale=0.5]{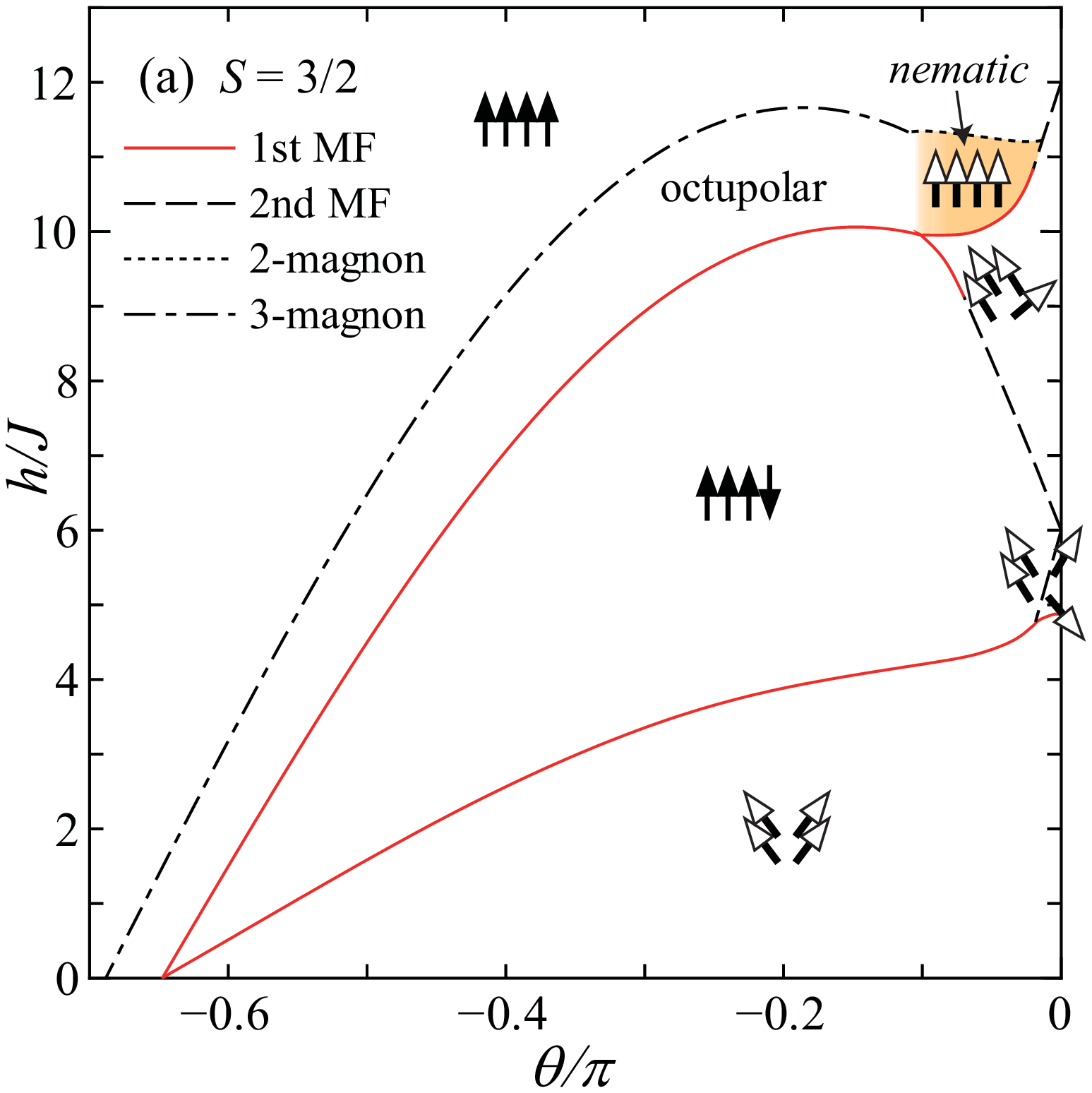}\\
  \includegraphics[scale=0.5]{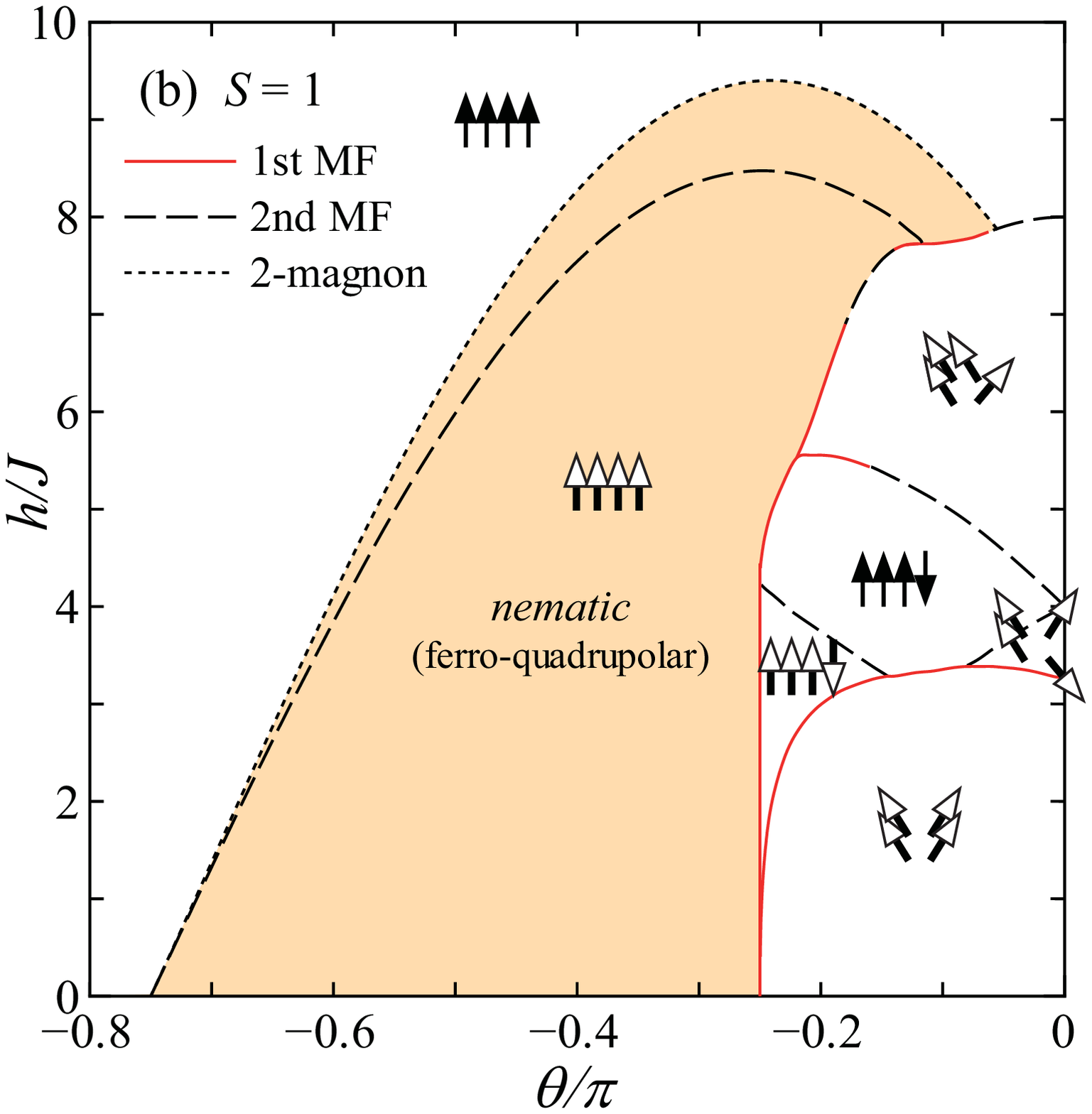}
  \caption{(Color online) Magnetic phase diagrams of the $S=3/2$ (a) and $S=1$ (b)
  pyrochlore antiferromagnets with the
  biquadratic interaction in a field $h$.
The symbols and notations are the same as Fig.~\ref{fig:pd}.
The dotted dashed line denotes the approximate boundary due to a three-magnon instability, which induces
an  octupolar phase.
The couplings are scaled as $J_1=J\cos \theta$
and $J_2=J\sin \theta$ with $J>0$.
  }
  \label{fig:pd2}
\end{center}
\end{figure}

Meanwhile in the MF 
approximation mentioned above, an octupolar state also
has the same energy as a quadrupolar state for
a finite $|J_2|$ at the saturation field.
The octupolar ordering can be regarded as a condensation of three-magnon bound states \cite{MomoiSS}.
To see how the degeneracy of the MF 
energies of quadrupolar and octupolar states is lifted, we also study the three-magnon excitations.
We calculated the energy of three-magnon bound states using a trial wave function.
We prepared on-site type bases, bond type bases, and plane type bases, which respectively contain
three magnons on the same sites, 
on the nearest neighbor two sites (bonds), and 
on the nearest neighbor three sites (planes).
Diagonalizing the $ 32 \times 32 $ matrix, we find that three-magnon bound states are also stable and have
a lower energy than three independent magnons for large negative $J_2$.
For small $|J_2|$, bound two-magnons are more stable than bound three-magnons,
but for large $|J_2|$, bound three-magnons become the most stable and the saturation field is
given by the instability line of bound three magnons [see Fig.\ \ref{fig:pd2}(a)].
In this approximation
the lowest bound three magnons have a flat mode, which is doubly degenerate for 
$ J_2<0 $, reflecting the existence of localized three-magnon bound states.
If we include more Hilbert space for three magnons or fluctuations around the degenerate ground states,
new quantum states for example an octupolar state might appear. This also remains to be a future problem.

Figure \ref{fig:pd2}(a) shows the obtained
whole phase diagram in a wider range of negative $J_2$,
in the notation $J_1=J\cos \theta$ and $J_2=J\sin \theta$ with
$J>0$.
We find that a small biquadratic interaction can arise magnon pairing in  $ S=3/2 $ pyrochlore antiferromagnets
and it gives the possibilities for the appearance of the magnon-pairing BEC phase, i.e.,
a spin nematic phase.

Lastly, we study the spin size dependence in the phase diagram, changing
the spins $S=3/2$ to $S=1$.
As we have discussed earlier, multipolar phases have been
found in $S=1$ spin systems on various lattices.
However, the $S=1$ bilinear-biquadratic model~\eqref{eq:j1-j2} on the
pyrochlore lattice has not been well investigated.
Following our study for $S=3/2$,
we obtained the phase diagram for $S=1$ in a similar manner, with
the MF approximation and the magnon instability analysis.
The obtained phase diagram is shown
in Fig.~\ref{fig:pd2}(b). 
It clearly shows that the region of the nematic phase is
much wider in the $S=1$ system than in the $S=3/2$ case.
Moreover, the nematic phase is enlarged, also
compared to the $S=1$ triangular lattice
bilinear-biquadratic model~\cite{Tsunetsugu,Lauchli2006}.
Therefore, if an $S=1$ pyrochlore Heisenberg antiferromagnet can be
synthesized, it would be a very good candidate for realization
of the spin nematic phase, even better than the $S=1$ triangular
lattice or the $S=3/2$ pyrochlore lattice.

To summarize, we investigated how quantum effects change magnetic phases
in pyrochlore antiferromagnets in applied magnetic field.
We find that a very small biquadratic interaction can induce magnon
pairing near the saturation, in $ S=3/2 $ pyrochlore
antiferromagnets. These magnon pairs give rise to a spin nematic phase
below the saturation field.  This spin nematic phase does not involve
any lattice distortion.
The characteristics of this state are
consistent with the results from ESR measurements
\cite{Hg_ESR}
and magneto-optical absorption in the novel phase
found experimentally in chromium spinel oxides
near the saturation field~\cite{Zn,Cd_new}. 
We hence expect it is indeed a realization of the spin nematic
phase.

It is our pleasure to acknowledge stimulating discussions with
Atsuhiko Miyata, Karlo Penc, Nic Shannon, and Shojiro Takeyama.
This work was supported by Grants-in-Aid for Scientific
Research (Grants Nos.~25103706, 15H02113, and 23540397)
from MEXT of Japan and by the RIKEN iTHES Project.
A part of this work was performed at the Aspen Center for Physics,
which is supported by US National Science Foundation grant
PHY-1066293.

\bibliographystyle{apsrev4-1}
\bibliography{spinel}

\end{document}